# Influence of pulse modulation frequency on helium RF atmospheric pressure plasma jet characteristics

Mahreen[#], G. Veda Prakash, Satyananda Kar*, Debaprasad Sahu and A. Ganguli

Department of Energy Science and Engineering, Indian Institute of Technology Delhi,
Hauz Khas, New Delhi – 110016, India

[#]mahreen.khanm@gmail.com, *satyananda@ces.iitd.ac.in

**Abstract**

This work investigates the influence of pulse modulation frequency ranging from 50 Hz – 10 kHz on the helium RF atmospheric pressure plasma jet's fundamental characteristics. The impact of modulation frequency on plasma jet discharge behavior, geometrical variation, reactive species emission, and plasma parameters (gas temperature $T_g$, electron excitation temperature $T_{exc}$, and electron density ($n_e$) are studied using various diagnostics such as optical imaging, emission spectra, and thermal diagnostics. From the experiments, it is observed that operating the plasma jet at low pulse modulation frequencies (around 50 Hz) provides enhanced plasma dimensions, higher electron densities and greater optical emission from reactive species (viz., He I, O, OH, $N_2^+$, etc.) as compared to the higher modulation frequencies. Besides the low power consumption, the three times less gas temperature of the modulated plasma jet than the continuous wave mode makes it more advantageous for the applications. Moreover, the influence of duty cycle (D) and applied RF power (P) on the plasma jet characteristics are also discussed. It is found that 10- 40% duty cycle operation provides the most favorable attributes. More importantly, the concern of shorter plasma length in RF plasma jets is overcome by operating at 10- 20% duty cycle with increased applied power. This work thoroughly characterizes helium atmospheric pressure RF plasma jet with a wide range of pulse mode operating parameters, which could help select appropriate operating conditions for various industrial and biomedical applications.

## I. Introduction

Low-temperature atmospheric-pressure plasmas have gained considerable attention in industrial and biomedical applications due to their enhanced reaction chemistry at near room temperature without the requirement of complex vacuum systems. [1–8] To understand and optimize the fundamental properties of these types of discharges, various plasma jet devices have been developed and studied in recent years [1,4,9–12]. Among them, the RF plasma jets have several advantages (over the low-frequency counterparts), like low breakdown voltage, high electron density, and more reactive species generation, etc. [6,13–16] Extensive studies are reported in this direction to understand the characteristics of RF plasma jet in terms of discharge modes, dynamics of charged particles, plasma parameters etc. [17–21] In atmospheric pressure RF discharges, the efficient power coupling enhances the considerable amount of gas temperature, which limits their application on heat-sensitive applications. [22,23] To overcome this limitation, researchers considered exciting the plasma discharge using RF pulse modulation, which controls the average power dissipation and reduces the gas temperature significantly. [22,24–26].

In this direction, several researchers have performed numerous experimental studies and simulations to explore the discharge behavior under RF pulse modulation. Balcon et al. have investigated the influence of pulse width on discharge operating mode, where the filamentary discharge was observed for pulse duration < 100 µs and glow type discharge for pulse duration > 1ms. [27] Another report conducted an experimental study of a pulse-modulated radio-frequency dielectric-barrier discharge in atmospheric helium. [28] By controlling the duty cycle (1-90 %) at a selective modulation frequency of 10 and 100 kHz, the 13.56 MHz discharge is shown to operate in three different glow modes: the continuum mode, the discrete mode, and the transition mode, and the discharge mechanisms were explained using residual electrons left in the electrode gap between two consecutive discharge events. Several studies were performed to explore further the influence of modulation variables on the plasma parameters and kinetics of charge particles and reactive species. Based on the experimental and simulated data, to improve the electron density and electron temperature in a pulse-modulated 13.56 MHz discharge, a lower modulation frequency, smaller than 50 kHz, with a duty cycle of 30-60 % was suggested by He et al. [25] Also, a 15 MHz RF power source at fixed pulse modulation frequency 100 kHz was simulated to obtain the electron density at variable duty cycles (20-100 %), and it was observed that the discharge characteristics on the duty cycle are sensitive in the region of around 40% duty cycle. [29] In these reports, the primary focus was on understanding the discharge transitions and basic plasma parameters at selective pulse parameters. In a recent report, controlled production of highly reactive atomic oxygen and nitrogen with a slight 0.1% air-like admixture ($N_2/O_2$ at 4: 1) for a fixed applied voltage, with a variation of duty cycle 20-100 % at 10 kHz selected pulsed radiofrequency (13.56 MHz) was demonstrated. [23] In another finding, He et al. presented a fluid model to investigate the generation of reactive oxygen species in an atmospheric pulse-modulated RF discharge with $He/O_2$ mixtures by varying the duty cycle from 30 to 60 % at a low modulation frequency < 50 kHz. [30] They also found that the electron density in a pulse-modulated RF discharge even becomes more significant than that in a continuous RF discharge at some duty cycles (30- 60 %).

With all these fundamental studies on atmospheric pressure RF plasmas, it was found that the effect of duty cycles on plasma properties is significant; therefore, similar studies are extended for pulse-modulated RF plasma jets. Platier et al. reported the utilization of microwave cavity resonance spectroscopy to estimate the electron density and effective electron collision frequency in the spatial afterglow of a 13.56 MHz pulsed driven atmospheric pressure helium plasma jet. [31] Modulation frequency was set between 125 and 8000Hz, and the duty cycle was varied from 40% to 0.625%, and during the "plasma on" phase, values of $1.7 \pm 0.3 \times 10^{18}$ m$^{-3}$ for the electron density and $0.12 \pm 0.01$ THz for the electron collision frequency were found. In another study, the effect of pulse modulation on the ionic content of a 13.56 MHz helium plasma needle was explored using molecular beam mass spectrometry. In all cases in their report, it was stated that the pulse repetition frequency was held constant at 6 kHz while the duty cycle was varied from 30% to 90%. [32] The results suggest that there is a possibility of tailoring the ionic plasma chemistry through pulsed modulation, as time-resolved ion intensity measurements demonstrated that negative ions were created almost exclusively in the pulse off-time while positive ions were generated mainly in the on-time. With the advancement of studies, the gas heating phenomenon is also investigated in the pulsed modulated RF plasma jet. Zhang et al. investigate the flow dynamics of an RF non-equilibrium argon atmospheric pressure plasma jet using the modulation frequency of 50 Hz (50 % duty cycle) and 20 kHz (20 % duty cycle). [14] Combined flow pattern visualizations (obtained by shadowgraph) and gas temperature distributions (obtained by Rayleigh scattering) are used to study the formation of transient vortex structures in the initial flow field. Apart from this, the application point of view of pulsed plasma jet was also studied, where a 13.3 MHz pulse-modulated plasma jet was operated at a fixed modulation frequency of 20 kHz with a 20% duty cycle. [24] The purpose was to experimentally investigate the plasma penetration and propagation mechanism inside capillary tubes with varying diameters for RF sustained helium and argon plasma jets. Plasma plume diameter in the free jet, 350 ± 80 μm and 675 ± 80 μm for the argon and helium jets respectively, was found significantly larger than the smallest tube size in which penetration was observed.

One notices that most of the studies on pulse-modulated plasma jets were performed either using a single or a couple of pulse modulation frequencies with a range of duty cycles. These studies demonstrated an additional control on the essential reactive oxygen and nitrogen species (RONS) and the heat transfer to the surface (with interest towards biomedical and heat-sensitive surface engineering) using RF pulse modulation parameters, specifically, duty cycle. [17,18,20,22,23]

Though the exploration of control on charge species and heat transfer using RF pulse modulation were made, the influence on fundamental characteristics viz., plasma shape (length and width) and discharge properties were rarely discussed. Extensive studies on the plasma jet dimensions (with control on heat transfer and species generation) and discharge dynamics are needed as the RF plasma jet suffers small dimensions. [33] Increasing the applied power to attain the longer plasma length led to thermal instabilities [34–36], whereas increasing the gas flow tends the plasma jet into turbulent operation. [37] This control on plasma jet dimensions could provide additional choice in transporting the generated species remotely to the targets where plasma-treated zone can be isolated from the plasma discharge region in various biomedical and material processing applications. [38]

Also, in the view of commercialization of RF cold plasma jets, a thorough knowledge of the influence of vast operating parameters on fundamental discharge characteristics (plasma dimensions, essential species characteristics, plasma parameters ($n_e$, $T_{exc}$, $T_g$) etc.) is highly needed. This has been considered to explore in the present work, especially for a sequential range of low-frequency pulse modulation regimes (<10 kHz) where minimal studies are available in the literature. It is worth mentioning that a low modulation frequency study is requisite as most commercially available standard RF generators have modulation frequency features only up to a few tens of kHz.

This study aims to experimentally investigate the influence of RF pulse modulation frequency from 50 Hz - 10 kHz on helium plasma jets fundamental characteristics such as discharge behavior, plasma dimensions, generation of reactive species, and the basic plasma parameters ($T_{exc}$, $n_e$, and $T_g$). Additionally, the influence of duty cycle (D) and applied power (P) on the plasma characteristics are explored. To the best of our knowledge, the attributes of helium cold plasma jet driven by the low pulse modulation frequencies (< 10 kHz range) have not been reported previously. One may motivate the present study further by considering the following arguments. While the power P is an appropriate measure of RF strength for CW systems, for pulsed systems it is $E_p$, the energy per pulse that is more appropriate. One has, $E_p = P\, T_{on} = P\, D\, T = P\, D\, /\, f_p = P_{avg}\, /\, f_p$. Here, T (= 1 / $f_p$) is the pulse repetition time, $T_{on} = DT$ is the on duration of the pulse, and $P_{avg} = P\, D$ is the average power. Thus, while it is important to operate in pulsed mode to bring gas temperatures in the 'safe' range for the selected application, it is also necessary to ensure that $f_p$ *is as low as possible* (for a given application). The latter in turn would ensure as high an $E_p$ as is acceptable, which would result in *higher electron densities*. It is this that has been a major guiding principle behind the present work.

The present manuscript is arranged as follows. Section II provides the details of the experimental setup and diagnostics utilized in the study. Section III presents the results and discussion part that illustrates the discharge characteristics by varying pulse modulation frequency, duty cycle, and RF power. A summary is given in section IV.

## II. Experimental setup

The developed plasma jet reactor consists of a glass tube (outer diameter 5 mm, inner diameter 3 mm, and length 10 cm) and two electrodes arranged in a cross-field configuration. A copper rod of diameter 1.5 mm was placed concentrically in the glass tube and used as the powered electrode, and a copper strip of 3 mm width was wrapped around the glass tube (3 mm above the glass nozzle), which was used as the ground electrode. Figure 1 (a) shows the schematic of the experimental setup and diagnostics arrangement. More details on the experimental setup can be found elsewhere. [12] Helium gas was used as the discharge gas, and the flow rate was controlled through a mass flow controller (1.5 – 7.5 lpm flow range). Applying RF power at 13.56 MHz (both continuous wave (CW) and pulse-modulated (PM) modes) using a commercial RF power generator (Cesar 1310) to the central electrode in the presence of helium gas inside the glass tube ignites the discharge, which flows out into the ambient air as a plasma jet.

The diagnostic tools used for the experiments are listed as follows. The amplitude of the pulse voltage was measured using a Tektronix (P6015A) 1000:1 high voltage probe, which was connected between the RF power generator and matching network to measure the applied voltage profiles without disturbing the plasma (as indicated in Figure 1(a)). The voltage waveforms were recorded by a Lecroy (HDO6104) digital oscilloscope. A typical voltage profile for 50 Hz, 30% duty cycle RF pulse modulation is shown in Figure 1(b). To capture the images, a Phantom (VEO 410) high-framerate (5200 fps) camera equipped with a Nikon 24-85 mm focal lens was used. All the images were recorded with 2200 μs exposure time. The images were captured in the front view (along the x-axis) and in some cases in axial view (i.e., viewed from below the discharge along the -z-axis); the specific view used is mentioned at the respective places in the manuscript. The precise plasma dimensions were estimated using the MATLAB image analysis. The captured images were processed to calculate the number of pixels corresponding to the reference dimensions (i.e., tube outer diameter 5 mm). For estimating the plasma parameters, two different optical emission spectrometers were used. To obtain the electron excitation temperature ($T_{exc}$), a spectrum in 200-900 nm was obtained using Ocean Optics HR 4000, 0.9 nm resolution spectrometer coupled with 200 μm optical fiber cable. Whereas estimating the electron density ($n_e$) $H_\beta$ spectrum was obtained using high-resolution Horiba Zobin 1250 mm, 8 pm resolution. In both cases, an optical fibre cable was attached to the translational stage and positioned at 3 mm from the plasma jet radially to record the maximum plasma emission from the discharge. In the helium spectra, 706 nm emission represents the high energy electrons, and thus, the intensities of significant species were normalized to the He 706 nm. [39] The plasma temperature was recorded using an insulated thermocouple, with an accuracy of ± 2°C.

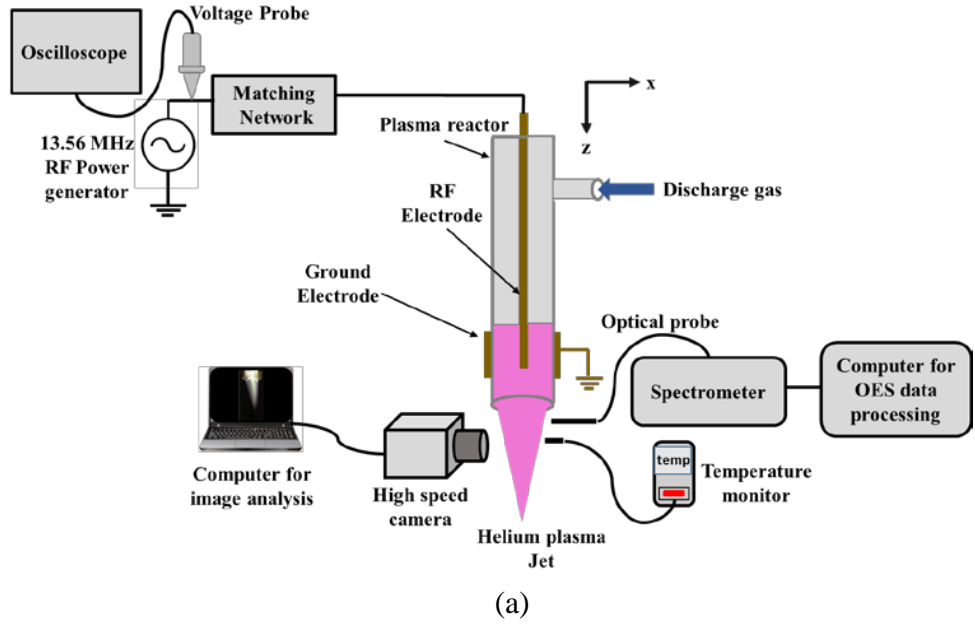

(a)

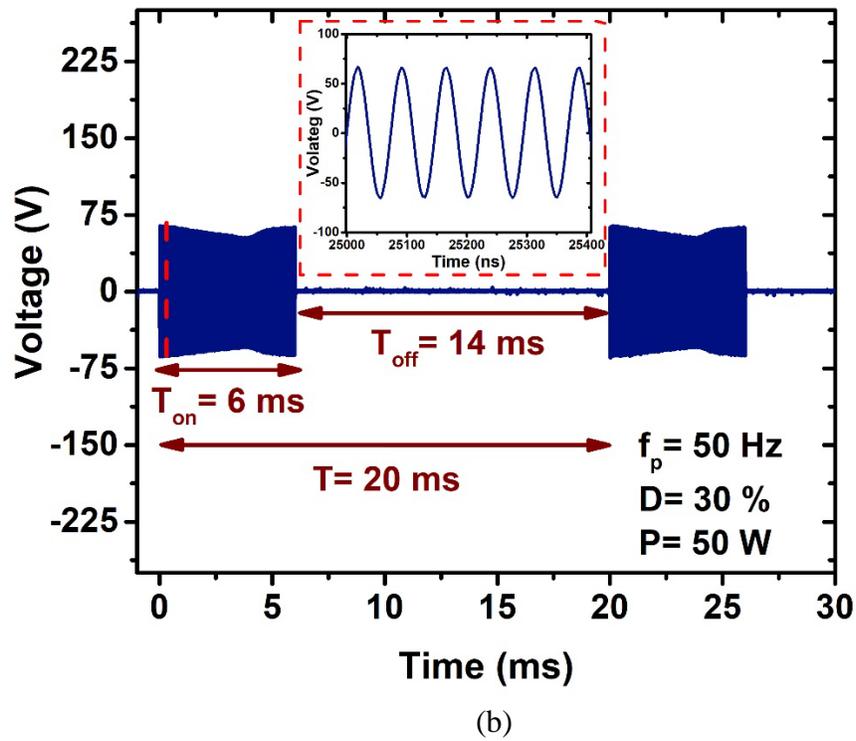

(b)

Figure 1: (a) Schematic of experimental setup and diagnostics arrangement, (b) typical voltage waveform of a 50 Hz, applied power 50 W, 30 % duty cycle RF pulse modulation.

## III. Results and discussion
### a. Effect of pulse modulation frequency on plasma jet discharge behaviour

The images of plasma jet recorded for CW and PM modes in the frequency range of 50 Hz - 10 kHz are shown in Figure 2 (a). The applied power (P = 50 W) and gas flow rate (Q = 1.5 lpm) are fixed for all cases. In Figure 2(a), the image for the CW plasma jet shows a more intense and radially contracted conical shape. The measured length of the plasma jet is ~ 7.45 mm, and the width just below the nozzle is ~ 2.25 mm. On the other hand, the images for pulse modulation frequency ($f_p$) from 50 Hz - 10 kHz (P= 50 W, Q= 1.5 lpm, and D= 30 % are fixed) show the plasma jet as weaker in intensity, smaller length, and more diffusive.

Specifically observing the case for 50 Hz modulation, the plasma jet shape is visible as a conical shape, shorter in length (~ 6.67 mm), broader (~ 2.69 mm), and homogeneous all over the volume with less intensity compared to the CW mode. With a further increment in the modulation frequency, the intensity, and the plasma jet length decreases. This reduction in length is related to less power dissipation in the discharge. At higher $f_p$, the small pulse duration limits the amount of power deposition to the plasma discharge, producing lesser ionization. Moreover, it is also observed that for frequencies above 2 kHz, the plasma jet is visible as discrete, split channels. This split in the plasma jet at a high frequency is related to the localized discharge formation on the edges of the inner electrode. At high frequencies, the smaller pulse duration limits the spreading of the discharge across the surface of the electrode (i.e., before the discharge can spread over the surface of the electrode, the applied power is turned off, extinguishing the discharge). More specifically as shown in section I, the *energy content* $E_p$ *of a single pulse of discharge is inversely proportional to* $f_p$, *i.e.,* $E_p = PD / f_p$. Thus, as the modulation frequency increases, $E_p$ decreases and a weak discharge is generated, whereas at low $f_p$, a relatively more intense and homogenous plasma jet results.

For example, at 50 Hz modulation frequency and 30% duty cycle, the total on time ($T_{on}$) is 6 ms, which corresponds to the total 81360 RF cycles fed to the discharge, whereas only 407 cycles were applied when the pulsing frequency increased to 10 kHz with 30 % duty cycle. The corresponding variation in the discharge intensity on the inner electrode is further shown in the images in Figure 2 (b) for 50Hz, and 10 kHz modulation frequencies. The images for 50 Hz show that the discharge is fully formed at the central electrode, whereas at 10 kHz the discharge is limited to the edges of the electrode, and the flowing gas is ionized in that limited region and propagates into the air as split channels.

Therefore, at fixed applied power, the number of RF cycles during the on phase is an essential factor determining the discharge behavior and the characteristics of the plasma jet. The dimensions of plasma jet length and width (at the nozzle and middle of the jet) as a function of modulation frequency are shown in Figure 2(c). It is seen clearly that the length of the plasma jet decreases as the pulse modulation frequency is increased from 50 Hz to 10 kHz. The plasma plume at 50 Hz modulation frequency, with input power of 50 W has a longer and broader shape and produces more homogeneous plasma with a plume length of (~ 6.5 mm). However, as the modulation frequency is increased to 2 kHz, the plasma plume becomes noticeably shorter with reduced width. With a further increment in the modulation frequency to 10 kHz, the plasma length reduces to ~ 4 mm. This reduction in plume length with increasing modulation frequency is due to progressively decreasing ionization on account of both lower *energy content* ($E_p$) deposited into the discharge as well as the *shorter time* available for ionization to occur efficiently. It may be noted that while the widths of the plume at the two locations (i.e., at the nozzle and middle of the plume) show a slight decrement up to modulation frequency 2 kHz, above 2 kHz, the plasma jet develops discrete channels that hinder exact measurement of the jet width. Overall, it is evident that the plasma jet shapes are dissimilar for these higher operating frequencies even though the input power is the same. As noted above, this is because the lower RF energy is deposited into the discharge with increasing modulation frequency, resulting in shorter plasma jet length and weaker plasma intensity. It is worth mentioning that increasing the applied power to 75 W results in increased plasma jet length and more homogenous plasma (i.e., without splitting of the plasma into discrete channels) even at 10 kHz modulation frequency, albeit with a higher plasma gas temperature. Such high applied power is not preferable in terms of power efficiency; thus, operating the RF plasma system at a lower modulation frequency (50 Hz) is suggested.

The gas temperature of the plasma jet is an important parameter and was measured for both CW and PM operating modes. An insulated K-type thermocouple was used to estimate the gas temperature of the plasma jet. Figure 2 (d) shows the variation of gas temperature as a function of applied power for both CW and PM modes. There is a significant temperature difference in the two modes of operation, which increases almost linearly with applied power. While the gas temperature reaches ~ 300 °C (at P = 50 W) in the CW mode, but in the PM mode, *the temperature tends to remain bunched within a small range* ~ 65 – 130 °C, as $f_p$ is reduced from 10 kHz to 50 Hz. What is noticeable is that even for 50 Hz modulation, the gas temperature drops significantly to about 120 °C or so, which is a significant improvement from the CW case and is also not too different from those obtained using high as $f_p$ values. Overall, it shows that operating in PM can obtain a large degree of controllability of gas temperature. The outcomes for the different key features of the plasma jet (its length, homogeneity, temperature, etc.) suggest that for optimum plasma jet shape, gas temperature, and plume intensity, it is preferable to operate in the low-frequency regime of PM mode (i.e., around 50 Hz).

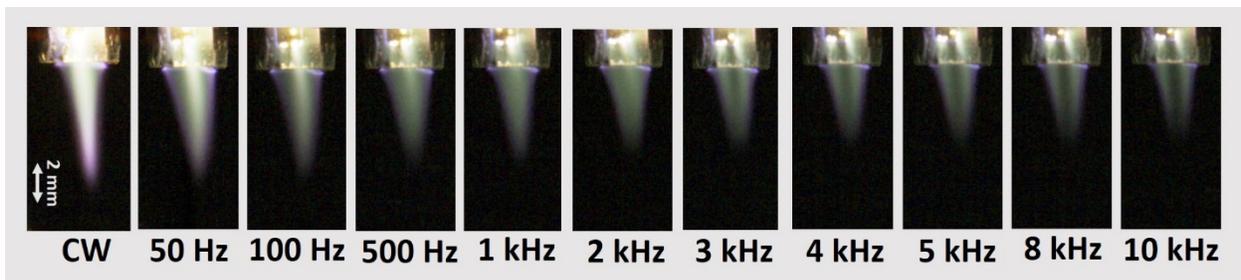

(a)

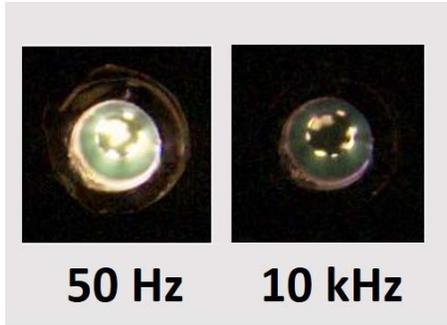

(b)

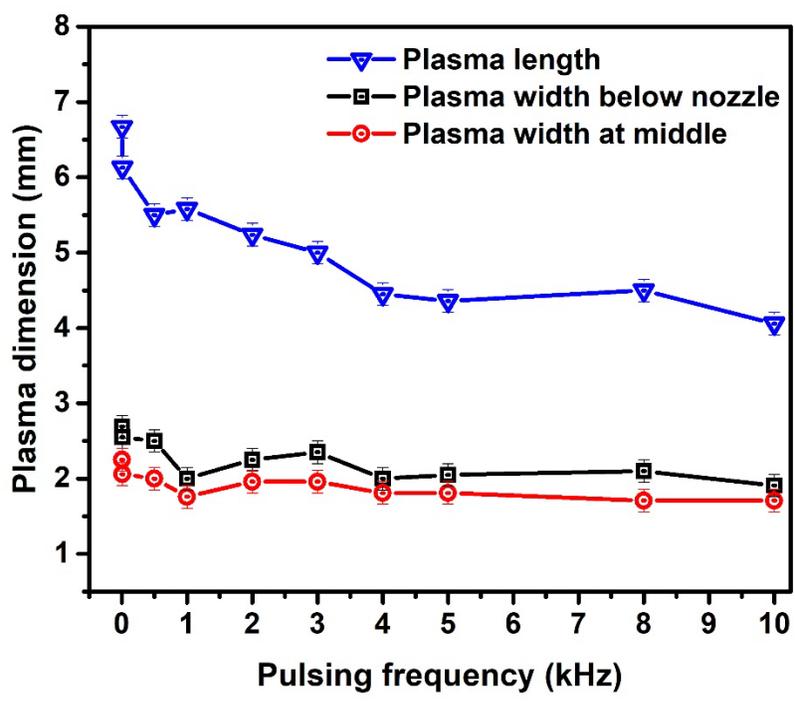

(c)

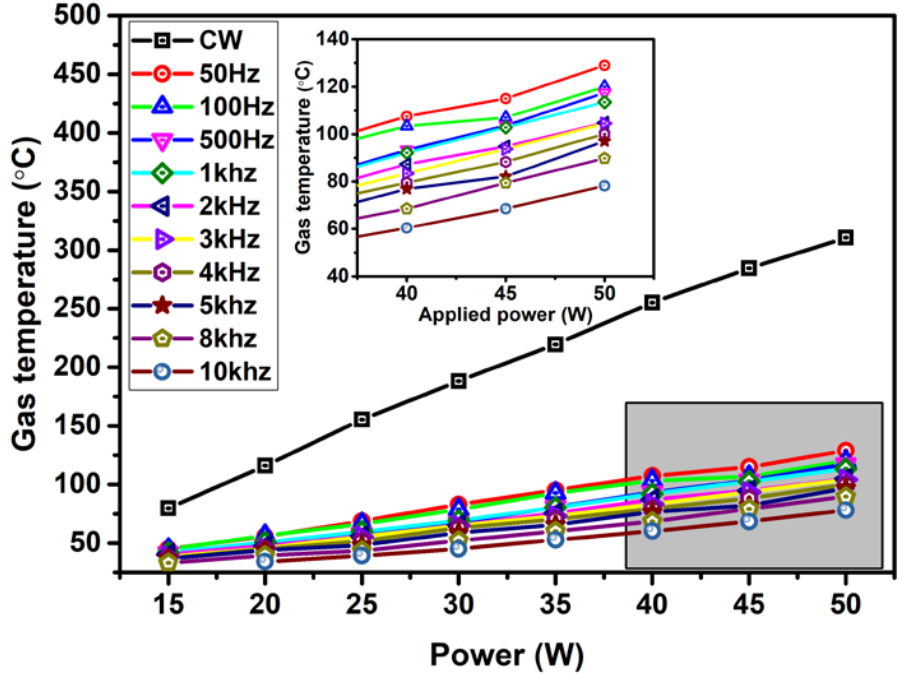

(d)

Figure 2: (a) Frontal images of a plasma jet in continuous wave (CW) and pulse mode (PM). (b) Discharge images for 50 Hz and 10 kHz plasma modulation frequency, captured axially from below the discharge, i.e., along the -z-axis (see Fig. 1(a)). (c) Length and width of plasma jet as a function of pulse modulation frequency (d) Plasma gas temperature variation with applied RF power for CW and different PM frequencies.

b. **Influence of pulse modulation frequency on the reactive species**

For many applications, apart from the plasma dimensions, the presence of various reactive species (OH, O, $N_2^+$, etc.) to participate in chemical reactions is important. To investigate the influence of pulse modulation frequency on the emission intensities from reactive species, the emission spectra of the plasma jet are obtained by using OES. The plasma emission is collected by placing the optical fiber cable at a 3 mm distance (radially) in front of the nozzle to collect the maximum emission intensity from plasma. The time-integrated optical emission spectra are recorded in the range 250–900 nm. Figure 3 (a) shows the spectral lines of helium plasma for operating parameters of $f_p$ = 50 Hz, P = 50 W, D = 30 % and Q= 1.5 lpm. The spectra show the emission of various important species (O, OH, $N_2^+$, etc.) along with the helium lines. Only the prominent emission lines in the spectra are identified using the NIST database [40].

In the plasmas, the energetic electrons produced during the discharge collide with the background helium atoms and stimulate the generation of excited and metastable helium atoms, which plays a vital role in initiating many chemical reactions in the plasma channel, viz.,

$$He + e \rightarrow He^* (2p) + e$$
$$He + e \rightarrow He^* (2S) + e$$

Helium atomic line of 706 nm [$3^3S \rightarrow 2^3P$], with the threshold excitation energy of 22.7 eV is an indicator of the presence of energetic electrons in the discharge and which is also the most intense spectral line in the helium plasma jet emission spectrum, shown in Figure 3(a). In comparison with the other He emission lines in the 200-900 nm wavelength range, 706 nm has the smallest excitation energy [41,42]; thus, intense emission occurs [42,43]. The other helium atomic lines noticed are 587.6 nm ($3^3D \rightarrow 2^3P$), 667.8 nm ($3^1D \rightarrow 2^1P$), and 728.1 nm ($3^1S \rightarrow 2^1P$). A helium atom naturally has two metastable levels, which are $2^1S_0$ (singlet) and $2^3S_1$ (triplet), with energies of 19.8 eV and 20.6 eV, respectively [44]. Helium metastable lines at He* 501.6 nm ($3^1P \rightarrow 2^1S$) and He*$_2$ 388.86 nm ($3^3P \rightarrow 2^3S$) are also observed in this work [42,44,45]. These highly energetic helium metastable atoms can ionize the nitrogen through the penning ionization during the interaction with the ambient air, producing reactive nitrogen species [46,47]. It is well known that nitrogen molecules effectively quench helium metastable states. Rapid penning reaction reduces the effective lifetime of helium metastable and produces excited nitrogen species. $N_2$ second positive system [$C^3\prod_u \rightarrow B^3\prod_g$] is observed in the spectrum (Figure 3 (a)) at the wavelength intervals of (315–381), in which the intense peaks are observed at the wavelengths of 337 (0–0), 357 (0–1), 375(1–3) and 380 nm (0–2). Two intense spectral lines of $N_2^+$ first negative system [$\Sigma_u^+ \rightarrow X^2\Sigma_g^+$] at the wavelengths of 391.44 nm and 427.8 nm are the other prominent productions of the discharge which are the indicators of helium metastable presence [42]. OH [$A^2\Sigma^+ \rightarrow X^2\pi\Pi$] at 308 nm are the other discharge productions present as the result of water vapor (present in the air) dissociation ($H_2O + e \rightarrow OH + H + e$), which is also labeled in Figure 3(a).

Most importantly, atomic oxygen intensity was found to be relatively higher among the RONS present in emission spectra, observed at $3p^5P \rightarrow 3s^5S$ (777.4 nm) and $3p^3P \rightarrow 3s^3S$ (844.6 nm). These relatively high emissions from the oxygen species were observed without admixing any external oxygen content to the main discharge gas, which is usually obtained by gas admixture. [30,41] Electron impact excitation of ground-state molecular and atomic oxygen leads to the emission at 844.6 nm and 777.4 nm, which is predicted by the dissociative excitation, given as,

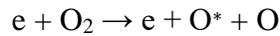
$$e + O_2 \rightarrow e + O^* + O$$

and direct impact excitation mechanisms,

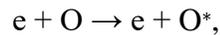
$$e + O \rightarrow e + O^*,$$

where O* refers to the O ($3p^5P$) state, which emits at 777.4 nm, and the O ($3p^3P$) state, which emits at 844.6 nm [48–50]. The generation of oxygen metastable is predominantly caused by electron impact dissociation of $O_2$, and electron-induced processes dominate the generation of ground-state atomic oxygen. The formation of such intense reactive species in helium discharge is beneficial for atmospheric pressure plasma applications, especially for biomedical treatment and plasma processing. [48–50]

Experimental observation of the important atomic and molecular emissions viz., He (706 nm), OH (309 nm), $N_2^+$ (391 nm), and O (777.4 nm) as a function of the applied modulation frequency is further explored and plotted in Figure 3 (b) - (e). From these figures, it can be observed that the emission intensity of the species is dependent on the applied pulsing frequencies. One notes that for lower modulation frequencies, the intensity of the different emissions (He, O, OH) in the plasma is greater. Figure 3 (b) shows the prominent helium lines at different modulation frequencies, which is because the higher modulation frequencies produce weak discharges. Thus, the production of high-energy electrons is reduced at higher modulation frequencies, which results in a decrease in the intensity of the helium line at 706 nm (also other helium lines). Therefore, the emission intensity of helium metastable is also decreased with increasing the modulation frequency and hence the reduction in penning ionization with nitrogen and oxygen. One can notice that the normalized intensities of the spectral lines of different RONS, namely OH, $N_2$, $N_2^+$, and O are significantly decreased with an increase in modulation frequency (Figure 3 (c-e)). For example, the normalized intensity of atomic oxygen was changed from ~0.55 to ~0.22, which shows a sensitive dependence on modulation frequency $f_P$. From this, it is evident that low modulation frequency operation provides relatively higher emissions from reactive species as compared to those obtained at higher modulation frequencies.

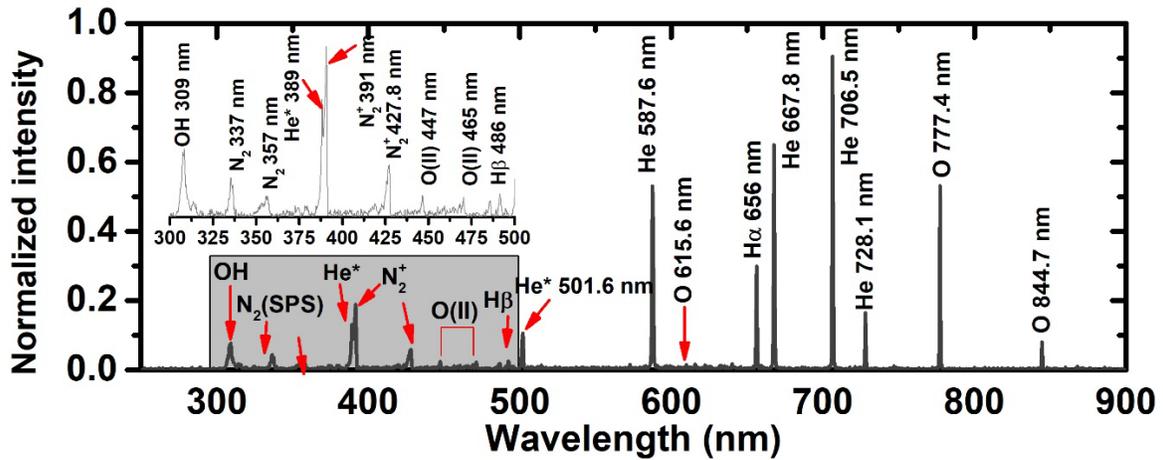

(a)

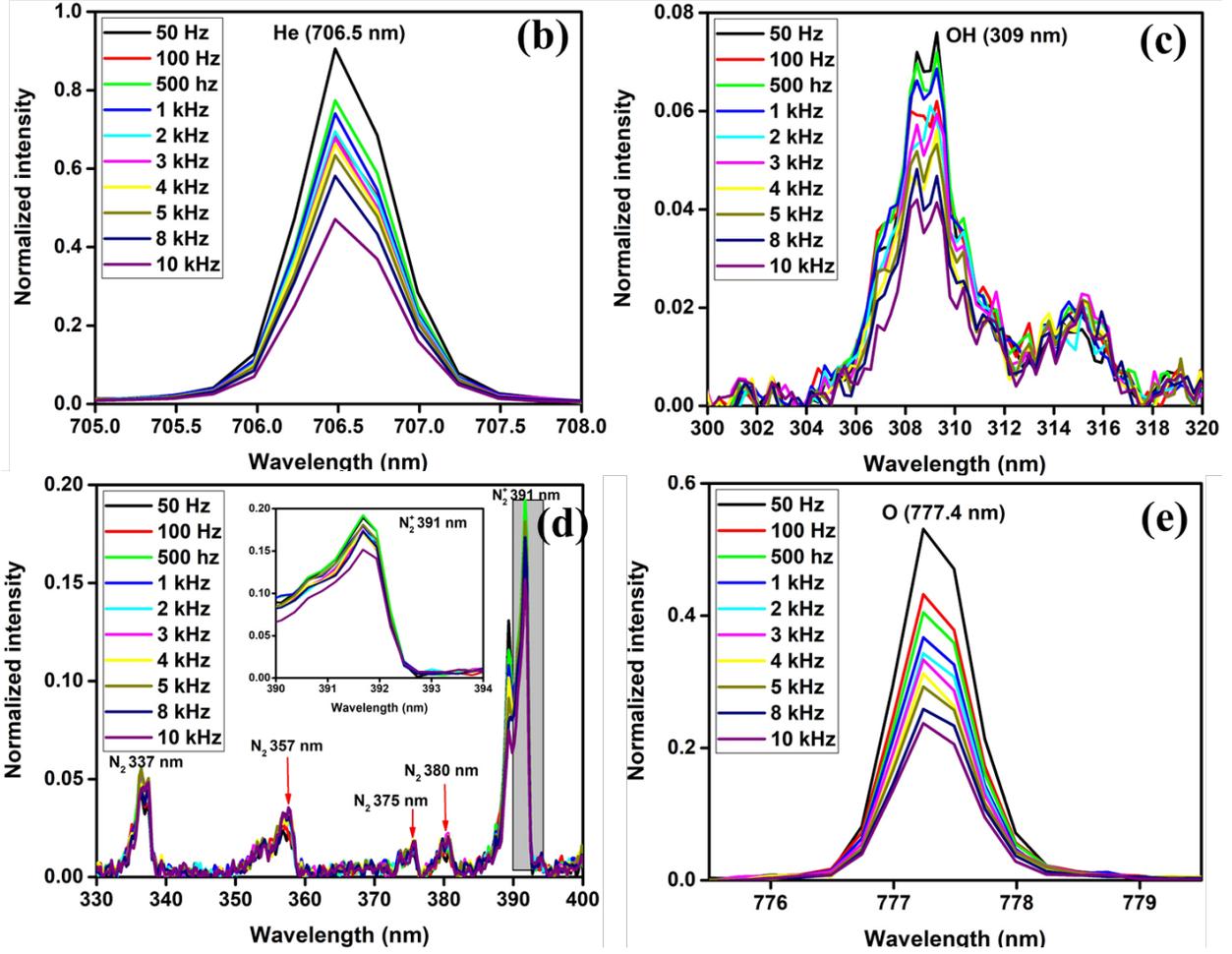

Figure 3: (a) OES of atmospheric pressure helium plasma jet in PM mode ($f_p$ = 50 Hz, P= 50 W, D= 30 % and Q= 1.5 lpm). (b) Emission spectra of helium line (706 nm), (c) OH line (309 nm), (d) $N_2$ and $N_2^+$ lines, (e) O line (777.4 nm) for various modulation frequencies.

### c. Influence of modulation frequency on plasma parameters

The basic plasma parameters such as electron excitation temperature ($T_{exc}$) and electron density ($n_e$) are estimated using the methods discussed in the reference. [16,51,52]

#### i. Estimation of electron excitation temperature ($T_{exc}$)

The electron excitation temperature is estimated using the Boltzmann plot method with the measured emission spectrum given in Figure 3(a). Spectral parameters of He I used in a Boltzmann plot to estimate $T_{exc}$ are given in Table 1 and a typical Boltzmann plot of helium spectra at 50 W applied power for 50 Hz modulation frequency is shown in Figure 4. The estimated $T_{exc}$ from the Boltzmann plot is found to be 1728 K.

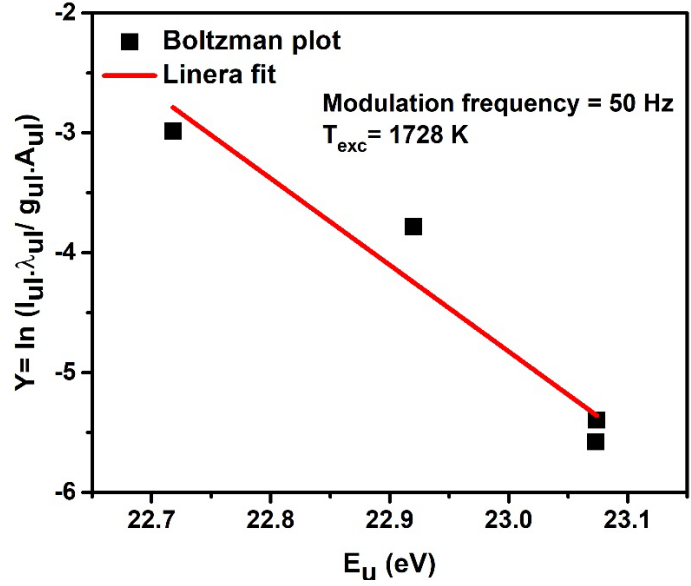

**Figure 4:** Typical Boltzmann plot fitting for electron excitation temperature ($T_{exc}$) estimation for 50 Hz, 50W, 30% duty cycle RF pulse modulation with 1.5 lpm gas flow.

Table 1: Spectral parameters of He (I) used in Boltzmann plot to estimate $T_{exc}$

| Wavelength (nm) | $A_{ul} \cdot g_{ul}$ | $E_u$ (eV) |
|---|---|---|
| 587.6 | $2.3 \times 10^8$ | 23.073 |
| 667.8 | $3.19 \times 10^8$ | 23.074 |
| 706.5 | $4.64 \times 10^7$ | 22.718 |
| 587.6 | $1.8 \times 10^7$ | 22.92 |

### ii. Electron density ($n_e$)

The electron density $n_e$ is one of the important plasma parameters which is estimated from the full width at half maximum (FWHM) of the $H_\beta$ spectrum (486.13 nm) using the Stark broadening method. The standard procedure was adopted from reference [16,51], where the $H_\beta$ spectrum is fitted with a Voigt profile to obtain the stark broadening ($\Delta\lambda s$). The electron density is determined using the following relation

$$n_e = \left(\frac{\Delta\lambda s}{2 * 10^{-15}}\right)$$

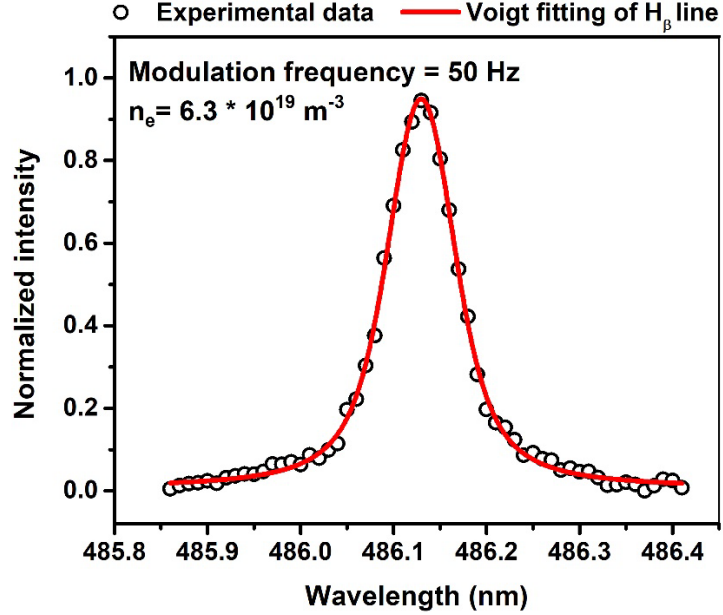

**Figure 5:** Normalized spectra line H$_\beta$ and corresponding Voigt-function fitting

Voigt fitting of experimentally measured H$_\beta$ spectrum at 50 W power, 50 Hz modulation frequency, and 1.5 lpm gas flow rate is shown in Figure 5. The estimated electron density is ~ $6.3 \times 10^{19} m^{-3}$. As the modulation frequency increases, the peak electron density decreases due to fewer RF cycles applied to deliver the power to the plasmas. Consequently, a modulation RF discharge with a lower modulation frequency is preferred to produce a higher electron density.

Electron excitation temperature and electron density as a function of modulation frequency obtained by the means above mentioned are shown in Figure 6. It is evident that the modulation frequency has an impact on both parameters. The electron excitation temperature is decreased up to 1 kHz modulation frequency above, which remains almost constant when the modulation frequency is increased from 50 Hz to 10 kHz, and the electron density is found to be decreasing in trend as the modulation frequency increases from 50 Hz to 10 kHz. The discrepancy in the $T_{exc}$ at a high modulation frequency (greater than 1 kHz) occurred due to the lower emission from the plasma and sensitivity of the OES technique. Therefore, the estimated values of plasma parameters at higher modulation frequency are ambiguous. However, in the presence of the evidence, optical imaging and species spectral lines indicate a diverse discharge mechanism involved at various modulation frequencies, and therefore, the electron excitation temperature and the density at 10 kHz must be small as compared to the low modulation frequency.

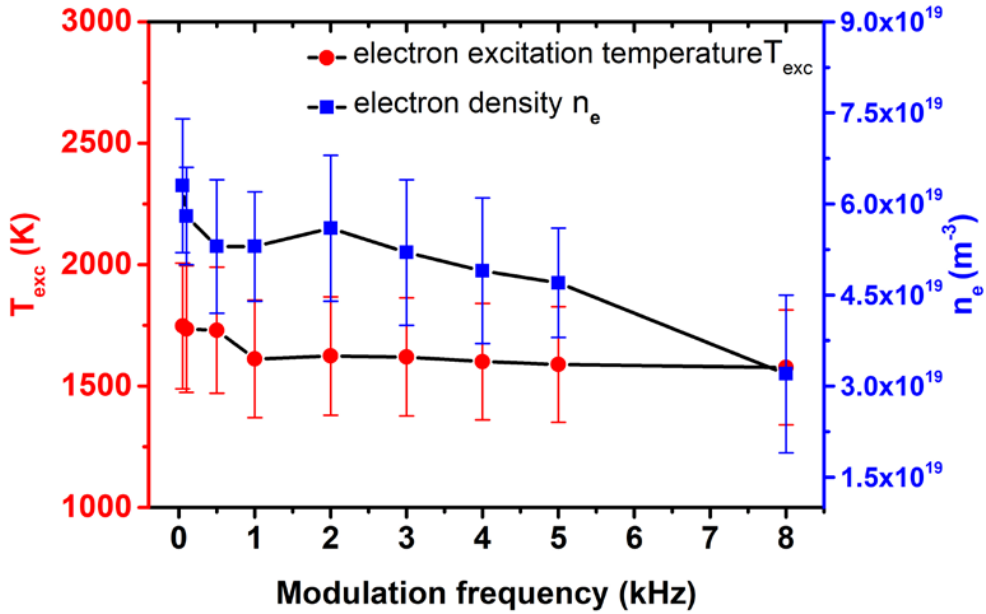

Figure 6: The variation of $T_{exc}$ and $n_e$ with reference to the modulation frequency at P = 50W, D = 30% and Q = 1.5 lpm He flow.

### d. Effect of fixed average power for CW and PM modes

To verify the influence of fixed average power for CW and PM modes, experiments are performed. The RF average power is fixed at 15W for CW and PM modes (50W applied power with 30% duty cycle corresponds to 15W average power). Corresponding images are given in Figure 7 (a), which clearly show that a weak discharge produces little extension into the ambient air in the CW mode, whereas the PM mode produces a long and broad plasma jet. This shows enhanced plasma jet dimensions over the CW mode, which could be an advantage for many practical applications of plasma jets. Similar observations were reported in the literature for microwave plasma jets [22,33]. Further, spectral analysis of the plasma in these two modes (shown in Figure 7(b)) illustrates the differences in their emission of atomic and molecular lines. Helium atomic emissions are higher in PM mode than those in CW mode, which indicates a greater number of high-energy electrons. Moreover, the intensities of the reactive species are comparable for both modes. For instance, the atomic oxygen (O) emission lines (777.4 nm and 844.6 nm) and OH have relatively higher emission intensity in the case of PM, whereas $N_2^+$ emission intensity is slightly higher in the case of CW. Therefore, the PM operation is beneficial in terms of lower power consumption and also provides additional control on essential reactive species intensities (such as O and OH) through the adjustment of input PM parameters. The control of these species is highly desired, especially for cancer treatment and wound healing. [8]

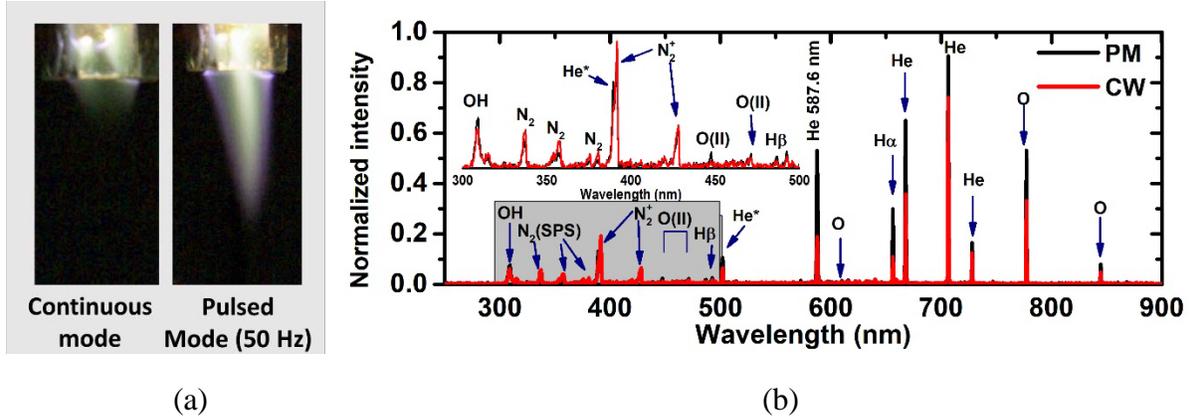

(a) (b)

Figure 7: (a) Images of a plasma jet in CW and PM modes at 15 W average power (D=30%, gas flow rate 1.5 lpm for PM mode). (b) Intensity comparison of plasma emission in CW and PM modes at 15 W average power.

### e. Effect of duty cycle

The duty cycle is one of the deciding parameters that control the amount of energy fed into the discharge. So, experiments were performed to observe the effect of duty cycle on plasma characteristics, as the latter was varied from 10 % to 80 %, with fixed P (= 50 W), Q (= 1.5lpm), and $f_p$ (= 50 Hz). The discharge becomes more intense and radially contracted for higher duty cycles (as it approaches the CW mode). However, there is no significant change in plasma length after the 40% duty cycle (Figure 8 (a)), as it attains the maximum plume length. From the images, even for duty cycle as low as 10%, adequate energy was fed to the discharge (duration for 10% duty cycle at $f_p$ = 50 Hz is 2 ms or 27120 cycles) so that the plasma attained sufficient ionization and hence adequate length yielding a low plasma-gas temperature of ~ 67.5 ℃.

In contrast, at higher $f_p$, there is a significant change in plasma jet shape and length observed. For example, in the images for 2 kHz modulation frequency shown in Figure 8 (b), varying the duty cycle from 10-80%, by keeping other parameters fixed (Q = 1.5 lpm and P = 50 W), at lower duty cycles (10 - 20 %), plasma appears to be weaker with split channels, and shorter in length indicating energy starvation during the pulse. The latter feature is offset by enhancement of the duty cycle, leading to more intense and homogeneous plasma and increased jet length, but radially contracted plasma at 80 % duty cycle. As already observed earlier, increment in duty cycle increases the number of RF cycles during the on period of the pulse, implying higher energy deposition and higher ionization that results in more homogeneous plasma with the extended length.

As mentioned at the beginning of Section I, the CW plasma jet is more contracted radially than the PM jet. Similarly, when the duty cycle is increased up to 80 %, the plasma jet contracts and tends to the CW mode of operation. This radial contraction of plasma is related to applying high duty cycle and high-power dissipation, which give rise to greater energetic electrons and increased gas temperature. In atmospheric pressure RF plasmas, the increase in dissipated power induces a higher current density, which leads to thermal instability due to the mode transition of the discharge from α to γ. [53–55] The gas ionization in the α mode (low current discharge supported by bulk plasma electrons) is volumetric throughout the electrode gap, whereas in the γ mode, the secondary electron emission strongly influences gas ionization which further increases the current density and leads high gas temperature and plasma contraction.[25,36,55] Therefore RF pulse-modulated discharge with adjusted duty cycle can generate stable plasmas reliably without the risk of mode transition.[36] Though the plasma ionization is increased at a higher duty cycle, producing more intense plasma with a long plume, the high gas temperature of the plasma makes it unsuitable for the treatment of heat-sensitive substrates. Therefore, it is suggested that the plasma jet can be operated at an optimum duty cycle, modulation frequency, and power, where the plasma has suitable gas temperature with the desired dimensions and can produce a sufficient amount of reactive species (RONS). From this parametric study, the reasonable choice of duty cycle is ~ 10- 40 %, with the modulation frequency adjusted suitably to obtain the desired plasma parameters.

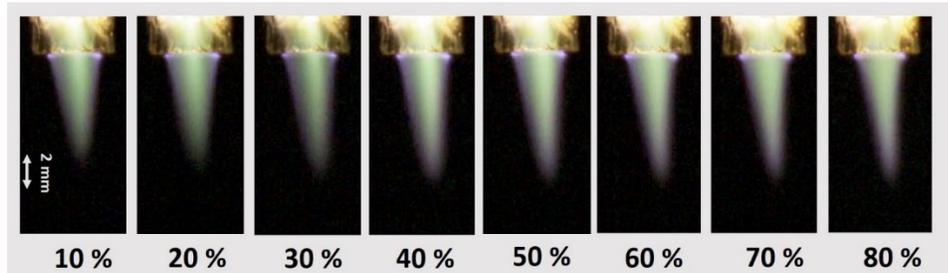

(a)

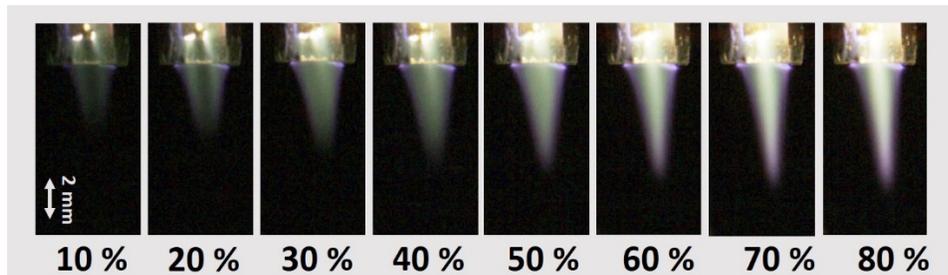

(b)

Figure 8: Images of helium plasma jet as a function of duty cycle (10 - 80%) at P= 50 W, Q= 1.5 lpm, (a) $f_p$= 50 Hz, (b) $f_p$= 2 kHz.

As discussed in Sec. III c, the influence of average power is found to be significant in CW and PM modes. Therefore, experiments were also performed to observe the impact of fixed average power (i.e., 15 W) in PM while varying the duty cycle (10 - 80 %). The corresponding applied powers at the different duty cycles are given in Table 1. It may be noted that at $D = 10\%$, instead of the required input power of 150 W, only 100 W was applied due to safety measures imposed on the RF generator, yielding an average power of 10%. For the other, the duty cycles of the applied power were as mentioned in Table 2. The maximum plasma jet length (~ 12 mm, shown in Figure 9) is observed at low duty cycles, mainly due to the high applied power. As the duty cycle increased from 10 - 80 % by keeping the average power fixed, the length of the plasma jet decreases gradually due to the *progressively lower amount of power applied* to the discharge.

The latter study clearly shows the need for a minimum power to sustain the desired discharge. Besides, operating the plasma jet at a lower duty cycle (10-20 %) with high applied power provides the solution of a shorter plasma plume which usually arises in helium RF plasma jet.[16,33] Acquiring a lengthy plasma jet was impossible in CW mode, as increased applied power leads to the discharge mode transition from glow to arc. This unique feature of the PM RF plasma jet offers more flexibility to the device for a specific application.

Table 2: Corresponding applied power for the duty cycle (10-80 %) for keeping the 15 W fixed average power

| Duty cycle $D$ (%) | Applied Power $P$ (W) |
|---|---|
| 10 | 150 |
| 20 | 75 |
| 30 | 50 |
| 40 | ~ 38 |
| 50 | 30 |
| 60 | 25 |
| 70 | ~ 22 |
| 80 | ~ 19 |

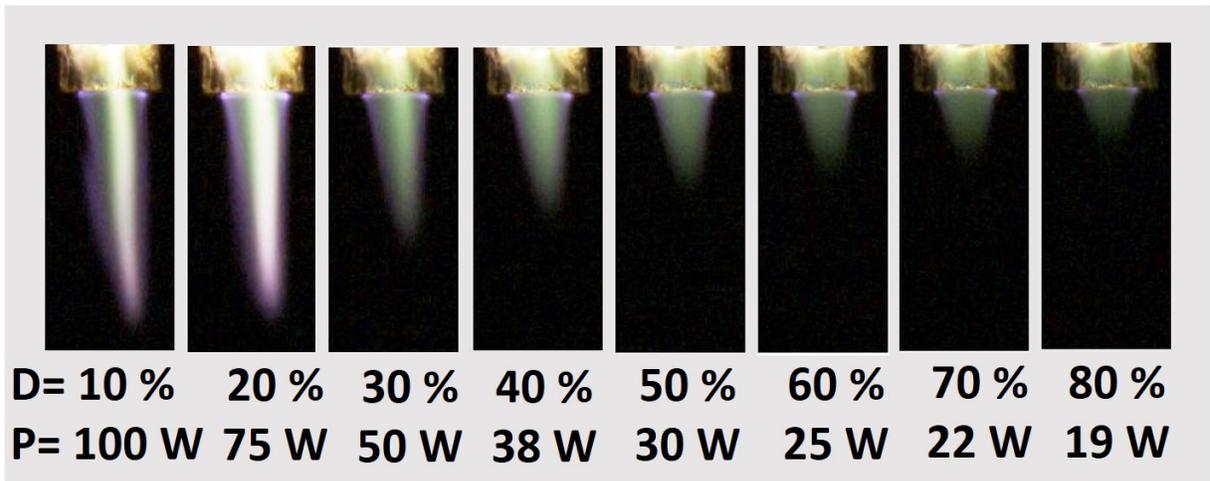

Figure 9: Images for fixed RF modulation power (15 W) with variation in duty cycle (10-80%) for $f_p$= 50Hz, gas flow 1.5 lpm.

### f. Effect of RF power

The applied RF power is the simplest and most common input parameter for controlling discharge properties and chemical reactivity of the plasma jets. Previous studies have shown that the plasma jet's length and its discharge properties can be controlled by varying the RF power. We have observed similar plasma behavior in terms of plume length, which increases along with the discharge intensity as the input RF power increases from 10 – 50 W, as shown in Figure 10.

The top row of images gives the front view, whereas the bottom row shows the discharge from the bottom (axial view) as the power increases from 10 W to 50 W. It is interesting to note that at low applied power, plasma ignites on the edges of the inner electrode (Figure 10, see the axial view, bottom row for 10W). This plasma discharge takes the jet shape as the applied power increases. Discrete, split channels are observed at low applied powers (up to 25 W) that gradually merge to form homogeneous plasma as the applied power is raised to 50 W, due to increased ionization. Plasma length is strongly dependent on the applied power, which increases with applied power, whereas the jet width decreases continuously when RF power is increased to 50 W.

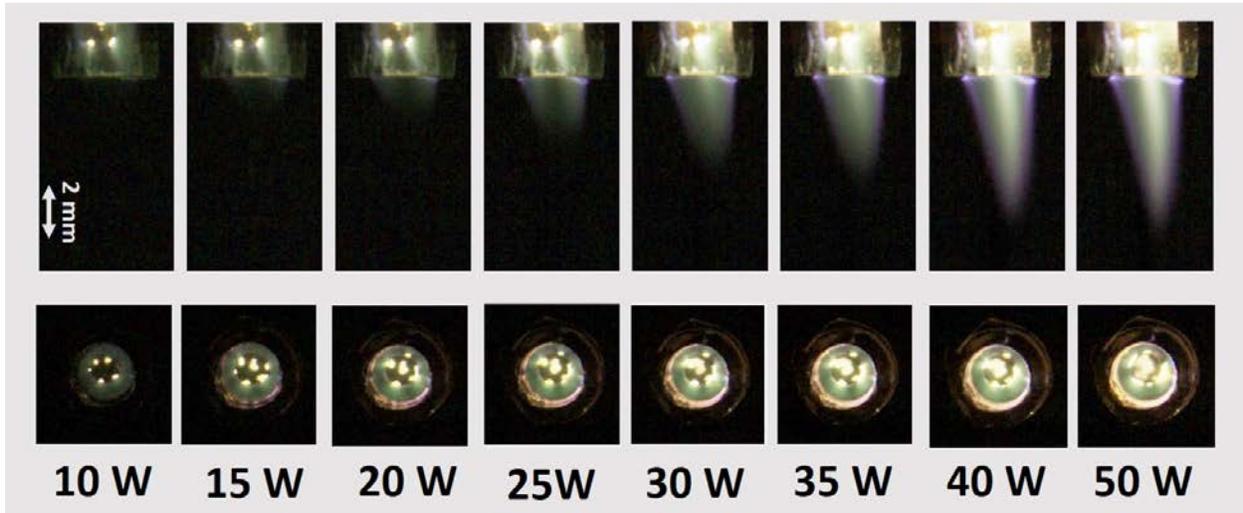

Figure 10: Images of helium plasma jet as a function of applied power at D= 30%, Q= 1.5 lpm, and $f_p$= 50 Hz (Top row: front view of the plasma jet, Bottom row: corresponding axial view (-z-axis)).

The relative variation of plasma dimensions as a function of duty cycle and applied power are plotted and shown in Figure 11 (a-b). The plasma jet length increases with the duty cycle and the applied power, whereas the plasma widths vary inversely. However, applied power controls the plasma length and width more than the duty cycle. Further, spectral analysis is also performed to note the influence of the duty cycle and applied power on the plasma species. Almost a similar trend of helium and reactive species intensities (OH, $N_2$, and O) is observed, as duty cycle and RF power are varied from 10- 80 % and 10- 50 W, respectively (Figure 11 (c-d)). The emission intensities are normalized in each case with respect to the maximum emission line of helium 706 nm. Specifically, the atomic oxygen (O 777.4 nm) line follows the trend of helium 706 nm, which explains the dependence of O generation on high-energy electrons.

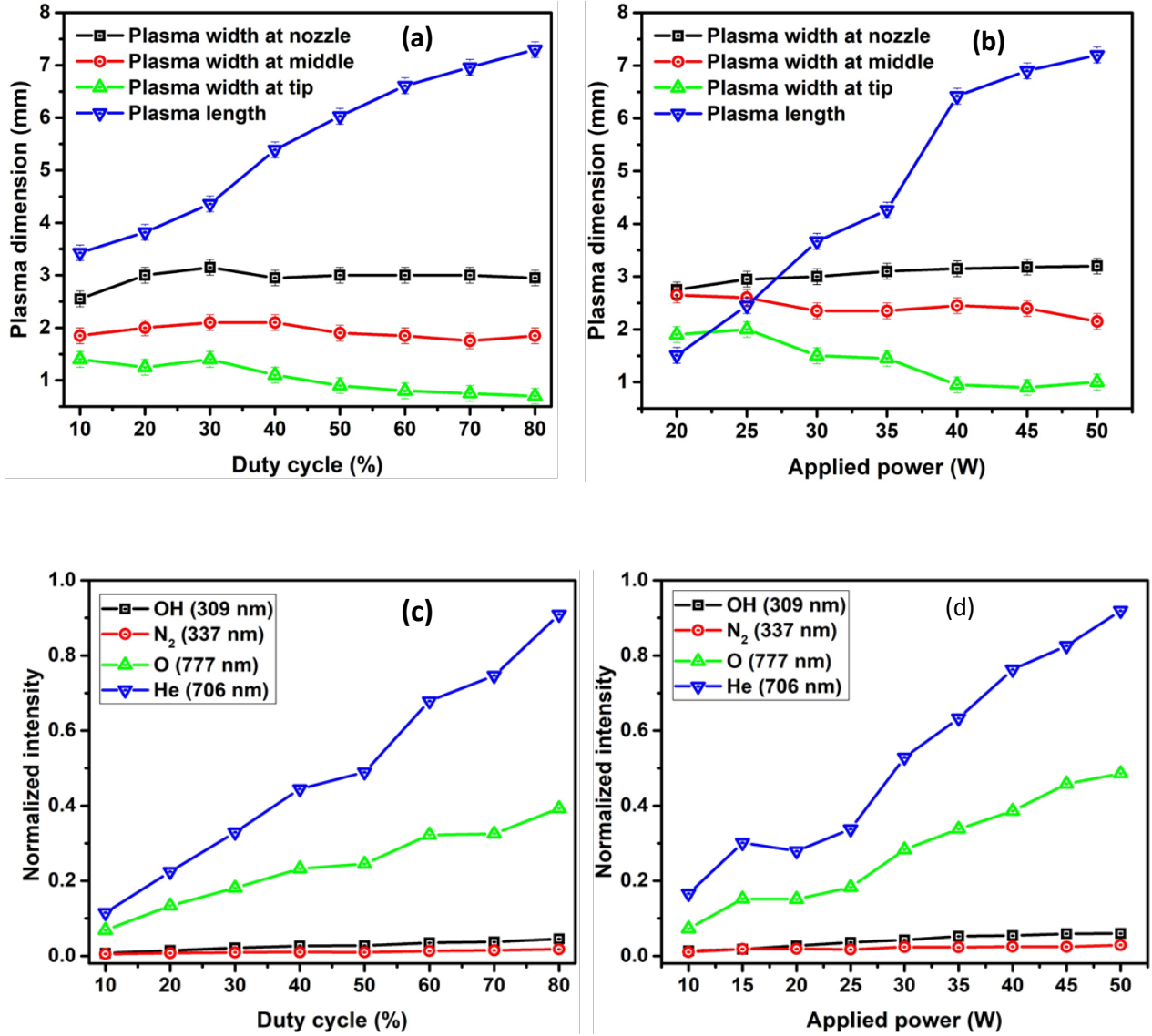

Figure 11: Plasma jet dimension variation at Q = 1.5 lpm for (a) duty cycle at P = 50 W and (b) applied RF power at D = 30%. RONS intensity variation for (c) duty cycle at P = 50W and (d) applied RF power at D = 30 %.

The above results suggest that for 50 W applied power and 30 % duty cycle, the low frequency, i.e., 50Hz pulse modulation, can generate a homogenous and long plasma jet with various essential reactive species. Whereas the high pulse modulation frequency (i.e., 10 kHz) generates a weaker and shorter plasma jet, an increase in applied power is required at a higher modulation frequency to develop the more intense and more prolonged plasma jet. Overall, it is found that the RF energy per pulse of discharge $E_p$, which depends on the RF power P, pulse modulation frequency $f_p$ and the duty cycle D has proven to be the most significant controlling factor that determines the plasma dimensions, emission intensities from reactive species, and gas temperature.

**IV. Summary**

1. The experimental outcomes for RF pulse modulation frequency ranging from 50 Hz – 10 kHz show that the optimum plasma jet characteristics (geometry, intensity, low gas temperature, and relatively high emission intensities from reactive species are obtained at low pulse modulation frequencies (where high $E_p$ prevails), i.e., around 50Hz (with ≤ 30% duty cycle) compared to high-frequency operation.
2. When the plasma jet is operated with the same average power in CW and PM modes, it is noticed that PM mode produces a better plasma jet, whereas in the CW plasma formation is weak and inadequate.
3. The duty cycle has no significant role on plasma dimensions and gas temperature at low modulation frequencies at around 50 Hz, whereas it has a substantial effect at higher modulation frequencies (>100 Hz). Moreover, a lengthy helium plasma jet can be generated in PM mode at a lower duty cycle by applying high RF power, which is impossible to develop in CW mode.
4. Applied power greatly influences the plasma jet length at all modulation frequencies. The number of RF cycles or duration of RF power and pulse modulation frequency has proven to be the significant control factors for plasma dimensions, reactive species formation, and gas temperature.

**Acknowledgements:**

The authors thank the reviewer for her/his careful evaluation of the paper that have contributed towards the significant improvement of the manuscript.

**DATA AVAILABILITY STATEMENT**
The data supporting this study's findings are available from the corresponding author upon reasonable request.